\documentclass[runningheads]{llncs}
\usepackage[T1]{fontenc}
\usepackage{booktabs} 
\usepackage{graphicx}
\usepackage{multirow}
\usepackage{orcidlink}
\vspace{-10mm}

\begin{document}
\title{
A Foundational EDM2-Based Generative Model for High-Resolution Synthetic Fetal Ultrasound Imaging from Open Datasets
}
%
%


%

\author{Harvey Mannering\inst{1}\orcidlink{0000-0003-3094-7417} \and
Yilin Zhang\inst{1}\orcidlink{0009-0008-3660-5737}  \and
Ziao Liu\inst{2}\orcidlink{0009-0006-1905-1061} \and
Zhiwu Huang \inst{1}\orcidlink{0000-0002-7385-079X} \and
Jacqueline Matthew\inst{3}\orcidlink{0000-0003-4754-0322} \and
Miguel Xochicale\inst{4}\orcidlink{0000-0002-8225-7517}}

\authorrunning{H. Mannering et al.}
\titlerunning{EDM2-Based Fetal Ultrasound Generation}
\institute{
$^1$ University of Southampton,
$^2$ Tsinghua University,
$^3$ King's College London, \\
$^4$ University College London
\thanks{
Work on diffusion models with the U. of Southampton supported by PhD studentship, Tsinghua University (classification), KCL (clinical evaluation) and reproducible workflows and robust software engineering on UCL-ARC’s Unified-AI cloud platform.
}
}

\maketitle              

\vspace{-5mm}
\begin{abstract}
Prenatal ultrasound imaging is key for assessing fetal health, but AI progress is limited by scarce, privacy-restricted, and hard-to-annotate datasets. 
We propose a high-resolution fetal ultrasound synthesis framework based on the EDM2 diffusion architecture, trained on multiple public datasets to generate 512×512 images across six anatomical classes. Our method achieved improved image quality with lower FID scores and enhanced downstream fetal plane classification, reaching 93.36\% ensemble accuracy after fine-tuning, surpassing real-data-only training. 
Clinical evaluation by an experienced fetal ultrasound specialist (10+ years) on 100 images yielded a mean realism score of 2.67/5, with real images rated higher than synthetic. Artefacts included smoothing, speckle irregularities, and anatomical inconsistencies. 
Code, data, models and other resources to reproduce this work are available at \url{https://github.com/xfetus/fetal-ultrasound-edm2}.
\end{abstract}

\vspace{-9mm}
\section{Introduction}
Prenatal ultrasound (US) imaging is the primary modality for assessing fetal health and development. Recent advances in AI have accelerated the development of classifiers, automated detection systems, and synthetic image generation methods for prenatal imaging. However, clinically reliable AI models require diverse datasets that capture real-world complexity, including varied fetal conditions, US scanners, patient demographics, and imaging modalities from 2D to emerging 3D-4D US \cite{kurjak2007useful}. Additional challenges include inconsistent volumetric assessment protocols, operator-dependent variability between sonographers, and natural anatomical variation between patients and fetuses \cite{jcm13185626}. 
Open datasets and evaluation frameworks are therefore essential for assessing AI model fidelity, robustness, and diversity \cite{IBRAHIM2025}.
Recently, \cite{Alsharid2025} published a comprehensive overview of publicly available US resources, including datasets and deep learning models, which is highly relevant to open-source strategies in US imaging. 
However, fetal US datasets remain largely centred on the FETAL PLANES DB, which contains 12,400 images \cite{burgos2020evaluation}. \cite{Alsharid2025} also reports additional emerging datasets, including the US Fetus Phantom dataset (15,728 images), the Fetal Head dataset (1,334 images), the Fetal Abdomen dataset (4,668 images), and the Fetal Cardiac dataset (300 images).
Parallel to this work, generative models such as GANs \cite{Goodfellow2014}, VAE \cite{kingma2013auto}, diffusion models \cite{ho2020_neurips}, and flow-based models \cite{lipman2022flow} are now a promising avenue to address data scarcity in medical images \cite{yan2018generation,tian2025enhancing}.
Diffusion-Based Fetal US Synthesis with Active Learning presented promising results on image quality comparison with baseline models but no clinical evaluation is included \cite{Arjemandi2026}. In US images, Stable Diffusion 1.5 \cite{rombach2022high} with ControlNet \cite{zhang2023adding} has enabled localized tumor synthesis \cite{freiche2025ultrasound}, while Tian et al. \cite{tian2025enhancing} utilized diffusion-generated images to improve plane classification, although only at $128 \times 128$ resolution. While hybrid Diffusion-GAN approaches \cite{iskandar2023towards} reached $256 \times 256$, these resolutions remain clinically restrictive and rely on older architectures. By contrast, we apply EDM2 \cite{karras2024analyzing}, the current state of the art for ImageNet generation \cite{deng2009imagenet}, to synthesize high-fidelity fetal US images at a more realistic $512 \times 512$ resolution.

\section{Methods and datasets}

We trained the EDM2 diffusion model to generate 6 US images classes with center cropping, random horizontal flipping, and resizing to $512 \times 512$, following the training setup from \cite{tian2025enhancing}, except with a higher resolution. We train two different sized networks, EDM2-S and EDM2-XL, which allows us to apply autoguidance \cite{karras2024guiding} to improve image quality.
We focus on the FETAL PLANES DB dataset (12,400 images) \cite{burgos2020evaluation}, which includes six classes: maternal cervix, fetal abdomen, fetal brain, fetal femur, fetal thorax, and other.
However, this dataset is relatively small, there is a risk of the diffusion model memorizing training data.
To mitigate this, we incorporate additional datasets: the FPU23 dataset (15,728 images) \cite{prabakaran2023fpus23}, a fetal abdominal structures segmentation dataset (1,588 images) \cite{da2023fetal}, and an African low-resource dataset (451 images) \cite{sendra2023generalisability}.
Each additional dataset is given a single label during training. 
We use a weighted MSE loss, weighting FETAL PLANES DB at 2.0 and others at 1.0.
Including these datasets allows twice as many training steps and reduces validation loss from 0.1427 to 0.1371.

\section{Experiments and results}

\begin{figure}[htbp]
\centering
  \caption{
  Representative fetal ultrasound images from real data, Tian et al. \cite{tian2025enhancing}, and our proposed high-resolution ($512 \times 512$) diffusion-based synthesis approach.
  Higher resolution image at our repository  \url{https://github.com/xfetus/fetal-ultrasound-edm2}.
  }
  \includegraphics[width=1.0\linewidth]{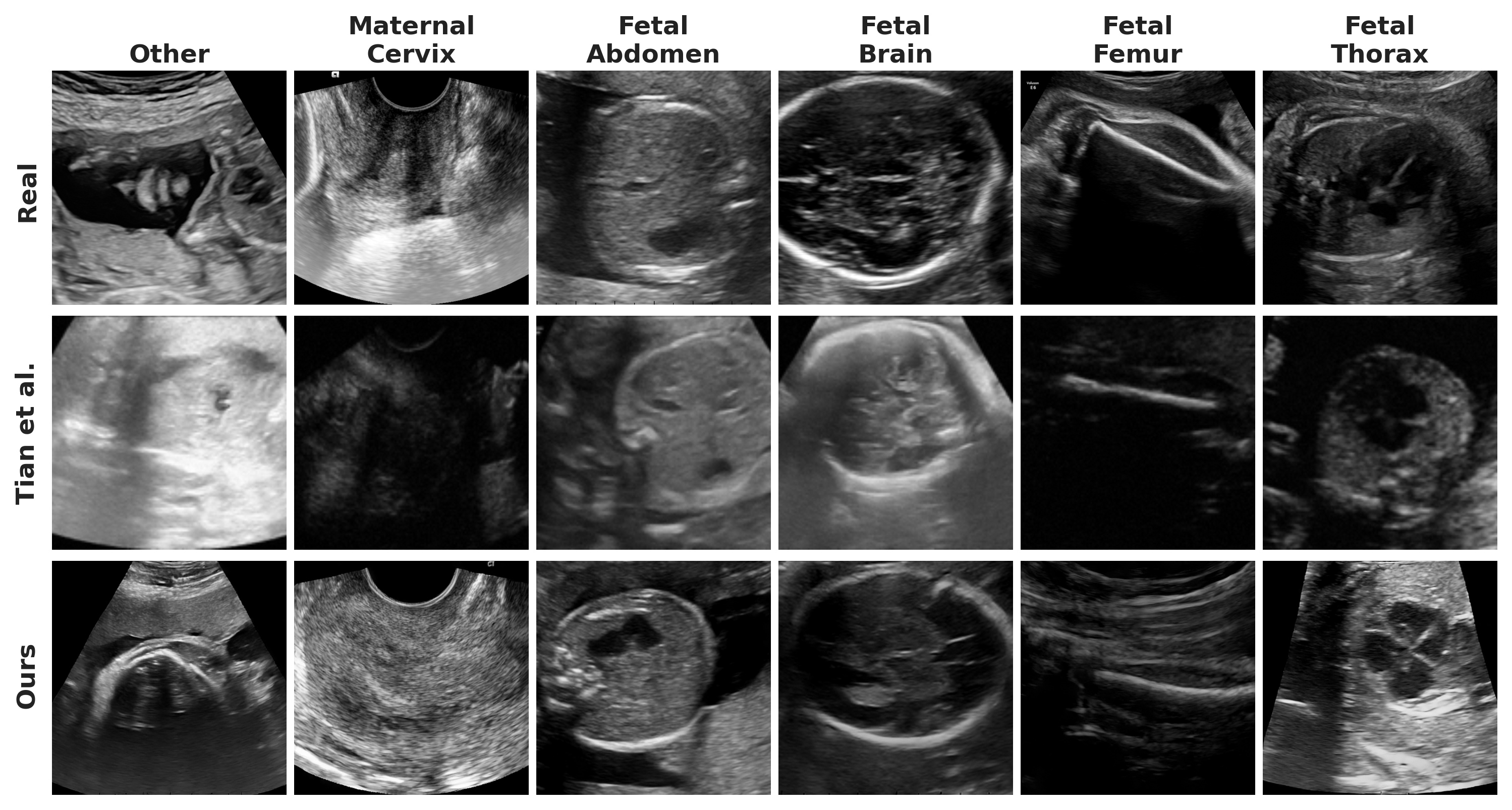}\label
  {fig:gen_images}

\vspace{-4mm}

\end{figure}

We evaluated the image quality of our image and compare them to \cite{tian2025enhancing} using Fréchet Inception Distance (FID) scores \cite{heusel2017gans}. We generate 5000 images for each of the six classes with a guidance scale of 2.25. For fair comparison, our images are downsized to 128 by 128. The results in Table \ref{tab:fid} show that our approach achieves higher quality in both individual classes and overall. Figure \ref{fig:gen_images} shows examples of generated images for each class.
Following Tian et al. \cite{tian2025enhancing}, who demonstrated that synthetic data improves fetal plane classification, we evaluated our high-resolution images using their top-performing architectures: ResNet50 \cite{he2016deep}, DenseNet169 \cite{huang2017densely}, and MedMamba \cite{yue2024medmamba}, plus a soft-voting ensemble. Using these specific models ensures a direct comparison with prior work.
We compared synthetic images generated by Tian et al. with our generated images under two settings: training from scratch on synthetic data only, and pretraining on synthetic data followed by real-world image fine-tuning. As shown in Table~\ref{tab:accuracy_comparison}, our images yielded stronger results when training from scratch, achieving 89.51\% ensemble accuracy. Under fine-tuning, our results also improved upon theirs, reaching a 93.36\% ensemble accuracy, surpassing the 92.32\% achieved on real-world data alone. These results suggest that our higher-resolution images are not only visually more realistic but also effective in supporting downstream classification performance.
An experienced fetal ultrasound clinician (10 years’ expertise) evaluated 100 generated images, distinguishing real from synthetic and rating quality on a 5-point Likert scale. The mean score was 2.67, with real images scoring higher (3.12) than synthetic ones (2.07). Judgement relied on subtle artefacts including smoothing, speckle patterns, and anatomical inconsistencies.

\vspace{-5mm}
\begin{table}[htbp]
  \centering
  
  \begin{minipage}{0.4\textwidth}
    \centering
    \caption{FID comparison between Tian et al. \cite{tian2025enhancing} and our generated images.
    Lower FID indicates better image quality.}
    \label{tab:fid}
    \scriptsize
    \begin{tabular}{lcc}
      \toprule
      \bfseries Class & \bfseries Tian et al.  \cite{tian2025enhancing} & \bfseries Ours \\
      \midrule
      Other & 189.14 & \bf 143.30 \\
      Maternal cervix & 269.08 & \bf 155.30 \\
      Fetal abdomen & 241.14 & \bf 150.12 \\
      Fetal brain & 184.06 & \bf 117.71 \\
      Fetal femur & 203.10 & \bf 141.57 \\
      Fetal thorax & 233.37 & \bf 135.99 \\
      Overall & 176.85 & \bf 104.25 \\
      \bottomrule
    \end{tabular}
  \end{minipage}
  \hfill 
  \begin{minipage}{0.56\textwidth}
    \centering
    \caption{Classifier accuracy comparison between Tian et al. \cite{tian2025enhancing} and our generated images.}
    \label{tab:accuracy_comparison}
    \scriptsize
    \begin{tabular}{llcc}
      \toprule
      \bfseries Method & \bfseries Model & \bfseries Tian et al. \cite{tian2025enhancing} & \bfseries Ours \\
      \midrule
      \multirow{4}{*}{\shortstack[l]{Training\\from\\Scratch}}
      & ResNet50     & 84.97\%  & \bf 85.94\% \\
      & DenseNet169  & 84.20\%  & \bf 87.67\% \\
      & MedMamba     & 82.53\%  & \bf 86.34\% \\
      & Soft Voting  & 87.35\%  & \bf 89.51\% \\
      \midrule
      \multirow{4}{*}{Fine-tuning}
      & ResNet50     & 91.29\% & \bf 92.45\% \\
      & DenseNet169  &  91.84\% & \bf 92.15\% \\
      & MedMamba     & 90.52\% & \bf 91.77\% \\
      & Soft Voting  & 92.84\% & \bf 93.36\% \\
      \bottomrule
    \end{tabular}
  \end{minipage}
\end{table}

\section{Conclusions and future work}
We present an EDM2-based foundational model for high-resolution fetal ultrasound synthesis using open datasets, improving image quality and downstream classification performance. The model generates 512×512 images, surpassing prior 256×256 approaches. Clinical evaluation of 100 images yielded a mean score of 2.67/5, with real images rated higher than synthetic outputs. All code and models are released, with future work targeting scalable foundation models for low-resource healthcare and further comparison of diffusion architectures.

\bibliographystyle{splncs04}

\begin{thebibliography}{10}
\providecommand{\url}[1]{\texttt{#1}}
\providecommand{\urlprefix}{URL }
\providecommand{\doi}[1]{https://doi.org/#1}

\bibitem{Alsharid2025}
Alsharid, M., Guo, X., Men, Q., Saha, P., Mishra, D., Ahuja, R., Ouyang, C.,
  Noble, J.A.: On the public dissemination and open sourcing of ultrasound
  resources, datasets and deep learning models. npj Digital Medicine
  \textbf{8}(1), ~777 (Nov 2025). \doi{10.1038/s41746-025-02162-4},
  \url{https://doi.org/10.1038/s41746-025-02162-4}

\bibitem{Arjemandi2026}
Arjemandi, M., Hassan, S., Wang, H., Valappil, S., Yaqub, M.: Difusal:
  Diffusion-based fetal ultrasound synthesis with active learning. In: Ni, D.,
  Noble, A., Huang, R., Xue, W. (eds.) Simplifying Medical Ultrasound. pp.
  130--139. Springer Nature Switzerland, Cham (2026)

\bibitem{burgos2020evaluation}
Burgos-Artizzu, X.P., Coronado-Guti{\'e}rrez, D., Valenzuela-Alcaraz, B.,
  Bonet-Carne, E., Eixarch, E., Crispi, F., Gratac{\'o}s, E.: Evaluation of
  deep convolutional neural networks for automatic classification of common
  maternal fetal ultrasound planes. Scientific Reports  \textbf{10}(1),  10200
  (2020)

\bibitem{da2023fetal}
Da~Correggio, K.S., Galluzzo, R.N., Santos, L.O., Barroso, F.S.M., Chaves,
  T.Z.L., Onofre, A.S.C., von Wangenheim, A.: Fetal abdominal structures
  segmentation dataset using ultrasonic images. Mendeley Data  \textbf{1}, ~1
  (2023)

\bibitem{deng2009imagenet}
Deng, J., Dong, W., Socher, R., Li, L.J., Li, K., Fei-Fei, L.: Imagenet: A
  large-scale hierarchical image database. In: 2009 IEEE conference on computer
  vision and pattern recognition. pp. 248--255. Ieee (2009)

\bibitem{freiche2025ultrasound}
Freiche, B., El-Khoury, A., Nasiri-Sarvi, A., Hosseini, M.S., Garcia, D.,
  Basarab, A., Boily, M., Rivaz, H.: Ultrasound image generation using latent
  diffusion models. In: Medical Imaging 2025: Ultrasonic Imaging and
  Tomography. vol. 13412, pp. 287--292. SPIE (2025)

\bibitem{Goodfellow2014}
Goodfellow, I., Pouget-Abadie, J., Mirza, M., Xu, B., Warde-Farley, D., Ozair,
  S., Courville, A., Bengio, Y.: Generative adversarial nets. Advances in
  Neural Information Processing Systems  \textbf{27} (2014),
  \url{https://proceedings.neurips.cc/paper/2014/file/5ca3e9b122f61f8f06494c97b1afccf3-Paper.pdf}

\bibitem{he2016deep}
He, K., Zhang, X., Ren, S., Sun, J.: Deep residual learning for image
  recognition. In: Proceedings of the IEEE conference on computer vision and
  pattern recognition. pp. 770--778 (2016)

\bibitem{heusel2017gans}
Heusel, M., Ramsauer, H., Unterthiner, T., Nessler, B., Hochreiter, S.: Gans
  trained by a two time-scale update rule converge to a local nash equilibrium.
  Advances in neural information processing systems  \textbf{30} (2017)

\bibitem{ho2020_neurips}
Ho, J., Jain, A., Abbeel, P.: Denoising diffusion probabilistic models. In:
  Larochelle, H., Ranzato, M., Hadsell, R., Balcan, M., Lin, H. (eds.) Advances
  in Neural Information Processing Systems. vol.~33, pp. 6840--6851. Curran
  Associates, Inc. (2020)

\bibitem{huang2017densely}
Huang, G., Liu, Z., Van Der~Maaten, L., Weinberger, K.Q.: Densely connected
  convolutional networks. In: Proceedings of the IEEE conference on computer
  vision and pattern recognition. pp. 4700--4708 (2017)

\bibitem{IBRAHIM2025}
Ibrahim, M., Khalil, Y.A., Amirrajab, S., Sun, C., Breeuwer, M., Pluim, J.,
  Elen, B., Ertaylan, G., Dumontier, M.: Generative ai for synthetic data
  across multiple medical modalities: A systematic review of recent
  developments and challenges. Computers in Biology and Medicine  \textbf{189},
   109834 (2025). \doi{https://doi.org/10.1016/j.compbiomed.2025.109834},
  \url{https://www.sciencedirect.com/science/article/pii/S0010482525001842}

\bibitem{iskandar2023towards}
Iskandar, M., Mannering, H., Sun, Z., Matthew, J., Kerdegari, H., Peralta, L.,
  Xochicale, M.: Towards realistic ultrasound fetal brain imaging synthesis.
  arXiv preprint arXiv:2304.03941  (2023)

\bibitem{karras2024guiding}
Karras, T., Aittala, M., Kynk{\"a}{\"a}nniemi, T., Lehtinen, J., Aila, T.,
  Laine, S.: Guiding a diffusion model with a bad version of itself. Advances
  in Neural Information Processing Systems  \textbf{37},  52996--53021 (2024)

\bibitem{karras2024analyzing}
Karras, T., Aittala, M., Lehtinen, J., Hellsten, J., Aila, T., Laine, S.:
  Analyzing and improving the training dynamics of diffusion models. In:
  Proceedings of the IEEE/CVF conference on computer vision and pattern
  recognition. pp. 24174--24184 (2024)

\bibitem{kingma2013auto}
Kingma, D.P., Welling, M.: Auto-encoding variational bayes. arXiv preprint
  arXiv:1312.6114  (2013)

\bibitem{kurjak2007useful}
Kurjak, A., Miskovic, B., Andonotopo, W., Stanojevic, M., Azumendi, G., Vrcic,
  H.: How useful is 3d and 4d ultrasound in perinatal medicine? Journal of
  perinatal medicine  \textbf{35}(1) (2007)

\bibitem{lipman2022flow}
Lipman, Y., Chen, R.T., Ben-Hamu, H., Nickel, M., Le, M.: Flow matching for
  generative modeling. arXiv preprint arXiv:2210.02747  (2022)

\bibitem{prabakaran2023fpus23}
Prabakaran, B.S., Hamelmann, P., Ostrowski, E., Shafique, M.: Fpus23: an
  ultrasound fetus phantom dataset with deep neural network evaluations for
  fetus orientations, fetal planes, and anatomical features. IEEE Access
  \textbf{11},  58308--58317 (2023)

\bibitem{rombach2022high}
Rombach, R., Blattmann, A., Lorenz, D., Esser, P., Ommer, B.: High-resolution
  image synthesis with latent diffusion models. In: Proceedings of the IEEE/CVF
  conference on computer vision and pattern recognition. pp. 10684--10695
  (2022)

\bibitem{sendra2023generalisability}
Sendra-Balcells, C., Campello, V.M., Torrents-Barrena, J., Ahmed, Y.A.,
  Elattar, M., Ohene-Botwe, B., Nyangulu, P., Stones, W., Ammar, M., Benamer,
  L.N., et~al.: Generalisability of fetal ultrasound deep learning models to
  low-resource imaging settings in five african countries. Scientific reports
  \textbf{13}(1), ~2728 (2023)

\bibitem{tian2025enhancing}
Tian, Y., Ucurum, E., Han, X., Young, R., Chatwin, C., Birch, P.: Enhancing
  fetal plane classification accuracy with data augmentation using diffusion
  models. IET Image Processing  \textbf{19}(1),  e70151 (2025)

\bibitem{jcm13185626}
Weichert, J., Scharf, J.L.: Advancements in artificial intelligence for fetal
  neurosonography: A comprehensive review. Journal of Clinical Medicine
  \textbf{13}(18) (2024). \doi{10.3390/jcm13185626},
  \url{https://www.mdpi.com/2077-0383/13/18/5626}

\bibitem{yan2018generation}
Yan, Y., Lee, H., Somer, E., Grau, V.: Generation of amyloid pet images via
  conditional adversarial training for predicting progression to alzheimer’s
  disease. In: International Workshop on PRedictive Intelligence In MEdicine.
  pp. 26--33. Springer (2018)

\bibitem{yue2024medmamba}
Yue, Y., Li, Z.: Medmamba: Vision mamba for medical image classification. arXiv
  preprint arXiv:2403.03849  (2024)

\bibitem{zhang2023adding}
Zhang, L., Rao, A., Agrawala, M.: Adding conditional control to text-to-image
  diffusion models. In: Proceedings of the IEEE/CVF international conference on
  computer vision. pp. 3836--3847 (2023)

\end{thebibliography}


\appendix

\section{Appendix A. Validation curves}

\begin{figure}[htbp]
\centering
\caption{
Validation loss curves for the UltrasoundEDM2 diffusion models. 
Larger model variants and additional training data consistently improve optimisation performance and final validation loss.
}
\includegraphics[width=1.0\linewidth]{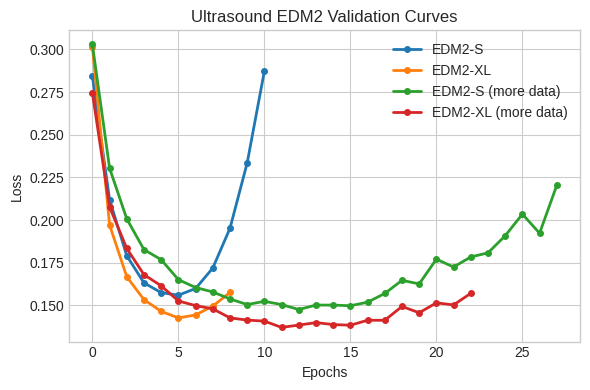}
\label{fig:val_curves}
\end{figure}

Figure~\ref{fig:val_curves} presents the validation loss curves for our UltrasoundEDM2 models on the FETAL PLANES DB dataset. 
As model capacity increases from the Small (S) to the Extra Large (XL) configuration, the optimal validation loss consistently improves, demonstrating the benefits of scaling the network architecture.

Training with additional datasets—including the FPU23 dataset, a fetal abdominal structures segmentation dataset, and an African low-resource ultrasound dataset—further improves performance. 
The increased diversity of training data enables the models to train for longer before convergence, ultimately achieving a lower validation loss than models trained on the FETAL PLANES DB alone.

\section{Appendix B. Ultrasound image evaluation survey}

\begin{figure}[htbp]
\centering
\caption{
GitHub Pages interface for the ultrasound image evaluation survey, available at \url{https://xfetus.github.io/fetal-ultrasound-edm2-survey-2026/}. 
The survey source code is publicly available at \url{https://github.com/xfetus/fetal-ultrasound-edm2-survey-2026}. 
The left panel shows the survey landing page, while the right panel illustrates the image evaluation interface used to answer the real-versus-synthetic classification and image quality assessment questions.
}
\includegraphics[width=1.0\linewidth]{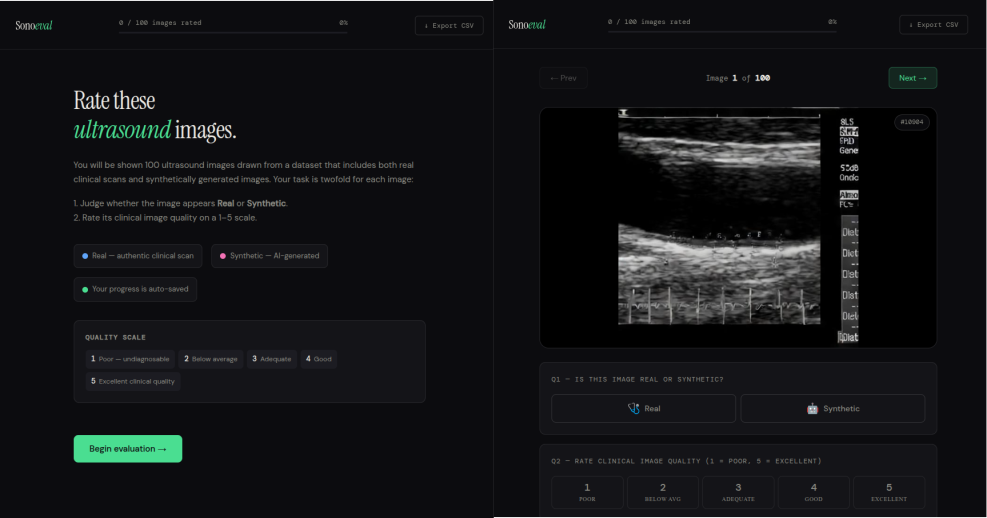}
\label{fig:survey}
\end{figure}

To support an independent assessment of image realism and clinical quality, we developed a lightweight web-based survey hosted using GitHub Pages. 
The survey enables clinicians and researchers to distinguish between real and synthetic ultrasound images while simultaneously rating overall image quality.
See Figure \ref{fig:survey} that illustrates landing page and image evaluation interface. 

\subsection{Survey design}

A fixed set of 100 images is sampled from the 30,000-image dataset using a deterministic random seed, ensuring that every participant evaluates the same images and enabling fully reproducible analysis.

Images are streamed directly from the Hugging Face Datasets Server API at runtime:

\url{https://huggingface.co/datasets/harveymannering/ultrasound_images_diffusion}

No local image hosting is required. Images are presented without labels and displayed in a randomised order to minimise potential bias.

\subsection{Evaluation protocol}

For each image, participants answer two questions:

\begin{itemize}
    \item \textbf{Q1.} Is the image \emph{Real} or \emph{Synthetic}?
    \item \textbf{Q2.} How would you rate the overall image quality on a five-point scale (Poor--Excellent)?
\end{itemize}

Once both responses have been recorded, the survey automatically advances to the next image. Progress is continuously saved in local browser storage, allowing participants to resume the survey after refreshing the page. Responses can be exported at any stage as a CSV file containing the image identifier, dataset index, real/synthetic classification, and quality rating.

\section{Appendix C. Open-access and reproducible research resources}

To promote transparency and reproducibility, all resources associated with this work are openly available. 
The project repository includes the source code, trained models, documentation, data processing scripts, and links to the accompanying preprint and related resources required to reproduce the experiments presented in this paper.

Project repository:

\url{https://github.com/xfetus/fetal-ultrasound-edm2}

Pre-trained model weights are publicly available via Hugging Face:

\url{https://huggingface.co/harveymannering/ultrasound-edm2}

\end{document}